\documentclass[a4paper, 12pt]{amsart}

\usepackage{amsmath, amssymb, amsthm}
\usepackage[english]{babel}
\usepackage[T1]{fontenc}

\newcommand{\be}{\begin{equation}}
\newcommand{\ee}{\end{equation}}
\newcommand{\bd}{\begin{displaymath}}
\newcommand{\ed}{\end{displaymath}}
\newcommand{\ba}{\begin{eqnarray}}
\newcommand{\ea}{\end{eqnarray}}

\def\R{{I \!\! R}}

\def\l{\lambda}

\def\v12{(v-w)}

\def\({\left(}
\def\){\right)}

\def\bgr#1\egr{{\allowdisplaybreaks\begin{gather}#1\end{gather}}}
\def\bma#1\ema{{\allowdisplaybreaks\begin{align}#1\end{align}}}

\def\oplem#1{\begin{lemma}\, {\rm #1}\, \it }
\def\cllem{\end{lemma}\rm \par }
\def\opthm#1{\begin{theorem}\, {\rm #1}\, \it }
\def\clthm{\end{theorem}\rm \par }

\def\N{\mathbb{N}}

\def\pRR{\hbox{{\tiny \rm I}\kern-.1em\hbox{{\tiny \rm R}}}}

\def\N{\mathbb{N}}
\def\R{\mathbb{R}}
\def\var{\varepsilon}

\def\var{\varepsilon}
\newcommand{\fer}[1]{(\ref{#1})}
\newcommand{\bq}{\begin{equation}}
\newcommand{\eq}{\end{equation}}
\def\bqa{\begin{eqnarray}}
\def\eqa{\end{eqnarray}}
\def\bd{\begin{displaymath}}
\def\ed{\end{displaymath}}

\newcommand{\dvv}{\frac{\partial^2}{\partial v^2}}
\newcommand{\dv}{\frac{\partial}{\partial v}}
\newcommand{\dt}{\frac{\partial}{\partial t}}

\newcommand{\dxi}{\frac{\partial}{\partial \xi}}
\newcommand{\dxixi}{\frac{\partial^2}{\partial \xi^2}}

\newcommand{\fs}{f_{\infty}}

\newcommand{\fc}{\widehat{f}}

\newcommand{\fq}{|\widehat{f}|^2}

\newcommand{\rd}{\rho_+}
\newcommand{\rs}{\rho_-}
\newcommand{\pp}{m_+}
\newcommand{\pn}{m_-}
\newcommand{\id}{\int_{0}^{+\infty}}
\newcommand{\is}{\int_{-\infty}^0}
\newcommand{\inr}{\int_{-\infty}^{+\infty}}
\newcommand{\Ha}{H_{\alpha}(f,\fs)}

\newtheorem{thm}{Theorem}
\newtheorem{cor}[thm]{Corollary}

\theoremstyle{remark}
\newtheorem{rem}[thm]{Remark}
\theoremstyle{definition}

\newenvironment{equations}{\equation\aligned}{\endaligned\endequation}

\begin{document}

\numberwithin{equation}{section}
\title{Wealth distribution in presence of debts. A Fokker--Planck description }

\author{M. Torregrossa}
\address{Department of Mathematics, University of Pavia,
via Ferrata 1,
Pavia, 27100 Italy}
\email{}
\author{G. Toscani}
\address{Department of Mathematics, University of Pavia, and IMATI CNR,
via Ferrata 1,
Pavia, 27100 Italy}
\email{}
\date{}

\maketitle

\maketitle
\noindent
{\bf Abstract:} \small{We consider here a Fokker--Planck equation with variable coefficient of diffusion which appears in the modeling of the wealth distribution in a multi-agent society. At difference with previous studies, to describe a society in which agents can have debts, we allow the wealth variable to be negative. It is shown that, even starting with debts, if the initial mean wealth is assumed positive, the solution of the Fokker--Planck equation is such that debts are absorbed in time, and a unique equilibrium density located in the positive part of the real axis will be reached.  }
\vskip 5mm


          \section{Introduction}\label{intro}

Mathematical modeling of wealth distribution has seen in recent years a remarkable development, mainly linked to the understanding of the mechanisms responsible of the formation of Pareto tails \cite{Par} (cf. Chapter 5 of \cite{PT13} for a recent survey). Among the various kinetic and mean field models considered so far \cite{Ch02, ChaCha00, ChChSt05, DMT, DMT1}, the Fokker--Planck type description of the evolution of the personal wealth revealed to be successful. In \cite{BM} Bouchaud and Mezard  introduced a simple model of economy, where the time evolution of wealth is described by an
equation capturing both exchange between individuals and random speculative trading,
in such a way that the fundamental symmetry of the economy under an arbitrary change
of monetary units is insured.  A Fokker--Planck type model was then derived  through a mean field limit procedure, with a solution becoming in time a Pareto (power-law) type distribution.
Let $f(v,t)$ denote the probability density at time $t \ge 0$ of agents with personal wealth $v\ge0$, departing from an initial density $f_0(v)$ with a mean value fixed equal to one
 \be\label{me1}
 m(f_0)= \int_{\R^+} v f_0(v) \, dv = 1.
 \ee
The evolution in time of the density $f(v,t)$ was described in \cite{BM} by the  Fokker--Planck equation 
 \be\label{FP2c}
 \frac{\partial f}{\partial t} = J(h) = \frac \sigma{2}\frac{\partial^2 }
{\partial v^2}\left( v^2 f\right) + \lambda \frac{\partial }{\partial v}\left(
(v-1) f\right),
\ee
where $\lambda$  and $\sigma$  denote two positive constants related to essential properties of the trade rules of the agents. 

The key features of equation \fer{FP2c} is  that, in presence of suitable boundary conditions at the point $v=0$, the solution is mass and momentum preserving, and approaches in time  a unique stationary solution of unit mass  \cite{TT1}. This stationary state is given by the (inverse)
 $\Gamma$-like distribution \cite{BM}
 \be\label{equi2}
f_\infty(v) =\frac{(\mu-1)^\mu}{\Gamma(\mu)}\frac{\exp\left(-\frac{\mu-1}{v}\right)}{v^{1+\mu}},
 \ee
  where the positive constant $\mu >1$ is given by
  $$ \mu = 1 + 2 \frac{\lambda}{\sigma}.
$$
As predicted by the observations of the Italian economist Vilfredo Pareto \cite{Par}, \fer{equi2} exhibits a power-law tail for large values of the wealth  variable.

The explicit form of the equilibrium density, which represents one of the main aspects linked to the validity of the model in its economic setting, is indeed very difficult to achieve at the Boltzmann kinetic level, where only few relatively simple models can be treated analytically \cite{BaTo, BaTo2,Kat}. 

In addition to \cite{BM}, the Fokker--Planck equation \fer{FP2c} appears  as limit of different kinetic models. It was obtained by one of the present authors with Cordier and Pareschi \cite{CoPaTo05} via an asymptotic procedure applied to a Boltzmann-type kinetic model for binary trading in presence of risks. Also, the same equation with a modified drift term appears when considering suitable asymptotics of Boltzmann-type equations for binary trading in presence of taxation \cite{Bisi}, in the case in which taxation is described by the redistribution operator introduced in \cite{BST}. Systems of Fokker--Planck equations of type \fer{FP2c} have been considered in \cite{DT2} to model wealth distribution in different countries which are coupled by mixed trading. Further, the operator $J(f)$ in equation \fer{FP2c} and its equilibrium kernel density have been considered in a nonhomogeneous setting to obtain Euler-type equations describing the joint evolution of wealth and propensity to trading \cite{DT1}, and to study the evolution of wealth in a society with agents using personal knowledge to trade \cite{PT-k}. 

These results contributed to retain that this limit model represents a quite satisfactory description of the time-evolution of wealth density towards a Pareto-type equilibrium in a trading society. 

Existence, uniqueness and asymptotic behavior of the solution to equation \fer{FP2c} have been recently addressed in \cite{TT1}. In this paper,  by resorting in part to the strategy outlined in \cite{FPTT16},  a precise relationship between the solution of the kinetic model considered in \cite{CoPaTo05} and the solution to the Fokker--Planck equation \fer{FP2c} was obtained, together with an exhaustive study of the large-time behavior of the latter. Various properties of the solution to equation \fer{FP2c} can in fact be extracted from the limiting relationship between the Fokker--Planck description and its kinetic level, given by  the bilinear Boltzmann-type equation introduced in \cite{CoPaTo05}. It is essential to remark that, in reason of the fact that the domain of the wealth variable $v$ takes values in $\R_+$, and that  the coefficient of diffusion depends on the wealth variable,  the analysis of the large-time behavior of the solution to equation \fer{FP2c} appears very different from the analogous one studied in \cite{AMTU, Tos99} for the classical Fokker--Planck equation.
In particular, the essential argument in \cite{TT1} was to resort to an inequality of Chernoff type \cite{C, Kla}, recently revisited in \cite{FPTT16}, that allows to prove convergence to equilibrium in various settings. 

All the previous results describe a society in which all agents have initially a non negative wealth, and do not consider the unpleasant but realistic possibility that part of the agents would have debts, clearly expressed by a negative wealth. 
Recent results on  one-dimensional kinetic models \cite{BP, BLM} showed however that there are no mathematical obstacles in considering the Boltzmann-type equation introduced in \cite{CoPaTo05} with initial values supported on the whole real line. 

Following the idea of \cite{BP, BLM}, we will study in this paper the initial value problem for the Fokker--Planck equation \fer{FP2c} posed on the whole real line $\R$, by assuming that the initial datum satisfies condition \fer{me1}, that is by assuming that part of the agents of the society could initially have debts, while the initial (conserved) mean wealth is positive. As we shall see, also in this situation, the positivity of the mean wealth will be enough to drive the solution towards the (unique) equilibrium density, still given by \fer{equi2}. Also, the forthcoming analysis will clearly indicate that the initial-boundary value problem considered in \cite{TT1}, in which the initial density is supported on the positive half-line, is simply a particular case of the general situation studied here. However, while the analysis of \cite{TT1} allows to conclude that the solution to the initial-boundary value problem for \fer{FP2c} converges strongly towards the equilibrium density \fer{equi2} with an explicit rate, in the general situation discussed in this paper, we are able to show that exponential in time convergence to equilibrium takes place only in a weak setting, well described by resorting to Fourier based metrics. 

As discussed in Section \ref{ent-mon}, the usual approach to convergence to equilibrium via entropy arguments fails in reason of the fact that in this situation the initial density and consequently the solution at each time $t>0$ is supported on the whole real line $\R$, while the equilibrium density is supported only on the positive half-line $\R_+$. This problem can  be bypassed by resorting to entropy functionals different from the standard relative Shannon entropy. 
However, a detailed evaluation of the entropy production of the new entropy functional allows to conclude only with a result convergence in the classical $L_1$ setting, without rate.

\section{Main results}\label{Bol}

\subsection{Existence and uniqueness}
Existence of a (unique) solution for the initial value problem for the Fokker--Planck equation can be recovered by means of the analysis done in \cite{TT1}, which is based on the strong connection between equation \fer{FP2c} and the kinetic equation of Bolzmann type introduced in \cite{CoPaTo05}. Indeed, the existence proof in \cite{TT1} is based on the Fourier transformed version of the kinetic equation, and applies without any change even if the wealth variable takes values on the whole real line. 
However, while Fokker--Planck equations with variable coefficients and in presence of boundary conditions have been rarely studied  \cite{Fel} (cf. also the book \cite{Feller} for a general view about boundary conditions for diffusion equations), in absence of boundaries, other results are available, which apply directly to the Fokker--Planck equation \fer{FP2c}. 

Particular  cases of Fokker--Planck type equations with variable coefficient of diffusion, mainly related to the linearization of fast diffusion equations have been studied in details (cf. \cite{CT07} and the references therein).  Then, the initial value problem for Fokker-Planck type equations with general coefficients has been recently investigated by Le Bris and Lions in \cite{LL}. Their results allow to conclude that the initial value problem for equation \fer{FP2c} has a unique solution for a large class of initial values.  In one-dimension of space Le Bris and Lions consider  Fokker-Planck equations in one of the the forms
\be\label{come5.2}
\dt p(v,t)= \frac{1}{2}\frac{\partial}{\partial v^2}\left({\sigma}^2(v) p(v,t)\right)+ \frac{\partial}{\partial v}\left( b(v)p(v,t)\right),
\ee
which corresponds to our case, equations in divergence form
\be\label{come5.8}
\dt p(v,t)= \frac{\partial}{\partial v}\left( \frac{1}{2}{\sigma}^2(v)\frac{\partial}{\partial v} p(v,t)+  b(v)p(v,t)\right),
\ee
and the so-called  backward Kolmogorov equation
\be\label{come5.3}
\dt p(v,t)= \frac{1}{2}{\sigma}^2(v)\frac{\partial}{\partial v^2} p(v,t)-b(v) \frac{\partial}{\partial v}p(v,t).
\ee
Let ${b}^{{\sigma}}$ and the Stratonovich drift ${b}^{s}$ be defined as
$$
\begin{aligned}
&{b}^{{\sigma}}={b}-\frac{1}{2}\frac{\partial}{\partial v}\sigma^2,
&{b}^{s}= b-\frac{1}{2}\sigma \frac{\partial}{\partial v}\sigma.
\end{aligned}
$$
Then, the following holds
\begin{thm}{\rm(\cite{LL})}\label{LL-cor2}
Let us assume that any one of the three drift functions ${b}$, ${b}^{{\sigma}}$ or ${b}^{Strat}$ satisfies
\be\label{LL-drift}
  {b(v)}\in W^{1,1}_{loc}(\mathbb{R}), \qquad \frac{\partial}{\partial v} {b(v)}\in L^\infty(\mathbb{R}), \qquad 
\frac{{b(v)}}{1+|v|}\in L^1+L^\infty(\mathbb{R}), 
\ee
and that ${\sigma}$ satisfies
\be\label{LL-sigma}
\sigma(v) \in W^{1,2}_{loc}(\mathbb{R}), \qquad \frac{{\sigma(v)}}{1+|v|}\in L^2+L^\infty(\mathbb{R}).
\ee
Then for each initial condition in $L^1\cap L^\infty(\mathbb{R})$ (resp. $L^2\cap L^\infty(\mathbb{R})$), the Fokker-Planck equation \eqref{come5.2}, the Fokker-Planck equation of divergence form \eqref{come5.8},  and the backward Kolmogorov equation \eqref{come5.3} all have a unique solution in the space
\be
 p\in L^\infty \big([0,T], L^1\cap L^\infty\big) \ \big(resp.\,\,\, L^\infty \big([0,T],L^2\cap L^\infty\big)\big),  \,\,\,
  {\sigma}\frac{\partial}{\partial v} p \in   L^2\big([0,T],L^2\big).
\ee
\end{thm}
The natural condition for the Fokker--Planck equation \fer{FP2c} is to apply Theorem \ref{LL-cor2} considering as initial value a probability density in $L^1\cap L^\infty(\mathbb{R})$. To this extent, it is sufficient to rewrite equation \fer{FP2c} in the divergence form 
\be
\dt f= \frac{1}{2}\dv \left(\sigma v^2 \dv f \right) + \dv \left[\bigg((\sigma+\l)v-\l \bigg) f \right], 
\ee
that is the analogous of equation \eqref{come5.8}, and to remark that in our case ${b(v)}=(\sigma+\l)v-\l  $ and ${\sigma(v)}=\sigma^{1/2} v$ .
\\
We obtain 
\medskip
\
\begin{thm}\label{exi}
Let $f_0(v)$  belong to $L^1\cap L^\infty(\mathbb{R})$. Then, the  the Fokker--Planck equation \fer{FP2c}, for $t \le T$, has a unique solution $f(v,t)$ in the space
\be
  f(v,t) \in L^\infty \big([0,T], L^1\cap L^\infty\big),   \,\,\,
  {v}\frac{\partial}{\partial v} f(v,t) \in   L^2\big([0,T],L^2\big).
\ee
\end{thm}

\subsection{Regularity}
The regularity of the solution to the initial-boundary value problem for equation \fer{FP2c} has been studied in \cite{TT1}. For the sake of completeness, and for its consequences on the large-time behavior of the solution, we give here a short proof.

For any given smooth function $\varphi(v)$, $v \in \R$ let us consider the weak form of equation \fer{FP2c}
\begin{equation}\label{wFP}
\frac{d}{dt} \inr \varphi(v) f (v,t) dv=  (\varphi, J(f))= \inr \left[\frac \sigma2  v^2 \varphi''(v) -\lambda (v-1)\varphi'(v)\right] \, f(v,t) \, dv. 
\end{equation}
Under the hypotheses of Theorem \ref{exi}, by
choosing $\varphi(v) = e^{-i\xi v} $ we obtain the Fourier transformed version of the Fokker--Planck equation \fer{FP2c}
\begin{equation}
\dt\widehat{f}(\xi,t)= \widehat J(\widehat f) = \frac{\sigma}{2}\xi^2 \dxixi \widehat{f}(\xi,t)-\lambda\xi \dxi\widehat{f}(\xi,t)-i\lambda\xi \widehat{f}(\xi,t) ,\label{fourif}
\end{equation}
where, as usual $ \widehat{g}(\xi) $ denotes the Fourier transform of $g(v)$, $v \in \R$
 \[
 \widehat{g}(\xi)  = \int_\R e^{-i\xi v} g(v)\, dv.
  \]
Let  $ \fc(\xi,t)=a(\xi,t)+ib(\xi,t)$. Then the real and imaginary parts of $\fc$  satisfy
\begin{equations}\label{rim}
\dt a(\xi,t)&=\frac{\sigma}{2}\xi^2 \dxixi a(\xi,t)-\lambda\xi \dxi a(\xi,t)+\lambda\xi  b(\xi,t),  \\
\dt b(\xi,t)&=\frac{\sigma}{2}\xi^2 \dxixi b(\xi,t)-\lambda\xi \dxi b(\xi,t) -\lambda\xi  a(\xi,t).
\end{equations}
Let us multiply equations \fer{rim} respectively by $2a$ and $2b$. Summing up we get the evolution equation satisfied by $|\fc(\xi,t)|^2$.
\begin{equation}
\dt \fq=\sigma\xi^2\bigg[ a \dxixi a+b \dxixi b\bigg]-\lambda\xi \frac{\partial \fq}{\partial \xi}. \label{eq-f2}\\
\end{equation}
Hence, multiplying by $|\xi|^p$ and integrating over $\mathbb{R}$ with respect to $\xi$, we obtain the evolution equation of the $\dot H_{p/2}-$norm of $f(v,t)$, where, as usual,  the homogeneous Sobolev space $\dot H_s$, is defined by the norm
 \[
 \|f\|_{\dot H_s}=\int_{\mathbb{R}}|\xi|^{2s}\, \fq(\xi)\,  d \xi.
 \]
We obtain
\begin{equation}
\dt \int_{\mathbb{R}}|\xi|^p  \,\fq\, d \xi=\sigma\int_{\mathbb{R}}|\xi|^{2+p}\bigg[ a \dxixi a+b \dxixi b\bigg]d \xi-\lambda\int_{\mathbb{R}} \xi|\xi|^p  \frac{\partial \fq}{\partial \xi}d \xi, \label{eq-HP}\\
\end{equation}
and integrating by parts the two integrals, it results
\begin{equation}
\dt \int_{\mathbb{R}}|\xi|^p \,\fq\, d \xi=(p+1)\bigg[\frac{\sigma}{2}(p+2)+\lambda\bigg]\int_{\mathbb{R}}|\xi|^p\fq d \xi-\sigma\int_{\mathbb{R}}|\xi|^{2+p} \bigg[ \big|\dxi a\big|^2+\big|\dxi b\big|^2\bigg] d \xi
\label{eq-HP2}.
\end{equation}
Since the last integral in \eqref{eq-HP2} can be bounded from below  \cite{TT1}
$$
\int_{\mathbb{R}}|\xi|^{2+p} \bigg[ \big|\dxi a\big|^2+\big|\dxi b\big|^2\bigg] d \xi\geq \frac{(p+1)^2}{4}\int_{\mathbb{R}}|\xi|^{p}\fq\, d\xi.
$$
we finally obtain
 \be
\dt \int_{\mathbb{R}} |\xi|^p \fq\, d \xi \leq \frac{p+1}2\bigg[ \sigma\frac{p+3}2 +2 \lambda \bigg]\int_{\mathbb{R}}|\xi|^p \fq d \xi.
\label{eq-HP3}
\end{equation}
The inequality \eqref{eq-HP3} implies that if the initial data has bounded $\dot H_p-$norm, then for all $t>0$, the $\dot H_p-$norm of the solution remains bounded, even if not uniformly bounded with respect to time. We proved
\medskip
\

\begin{thm}{\rm(\cite{TT1})}\label{reg}
Let $f_0(v)$ be a probability density in $\R$ that belongs to $\dot H_r(\R)$. Then, the $\dot H_r-$norm of the solution $f(v,t)$ to the Fokker--Planck equation \fer{FP2c}, for $t \le T$, still belongs to $\dot H_r(\R)$, and
 \be\label{gro-t}
\int_{\mathbb{R}} |\xi|^{2r} \fq(t)\, d \xi \leq  \exp \left\{\frac{2r+1}2\bigg[ \sigma\frac{2r+3}2 +2 \lambda \bigg]\,t \right\} \int_{\mathbb{R}} |\xi|^{2r} |\widehat f_0|^2\, d \xi.
\ee
\end{thm}
\medskip
\medskip
\begin{rem} The difficulty of recovering the uniform boundedness of the  $\dot H_r(\R)$-norm of the solution to the Fokker--Planck equation \fer{FP2c} is strictly related to the singularity of the coefficient of diffusion $\sigma v^2$, which vanishes in correspondence to the point $v=0$. Indeed, as proven in \cite{CT07} for a similar Fokker--Planck equation with coefficient of diffusion $1+\sigma v^2$, the uniform boundedness of the  $\dot H_r(\R)$-norm of the solution holds.
\end{rem}

\subsection{Further properties} 
The analysis of \cite{LL} do not care about the eventual preservation of positivity of the solutions to \fer{come5.2}. However, this property can be easily proved for equation \fer{FP2c}, by resorting to the same argument used in \cite{TT1} for the same equation posed in $\R_+$. Indeed, as proven in \cite{TT1}, the solution to the Fokker--Planck equation \fer{FP2c} is the limit of the solution to a kinetic equation of Boltzmann type, for which it is elementary to obtain the positivity property. 

Positivity can however be proven directly by working on the Fokker--Planck equation, 
by resorting to the following argument \cite{Ill}.
Suppose the initial data $f_0(v)$ (and hence the unique solution) to the Fokker--Planck equation \fer{FP2c}  are smooth and vanish  for $v= \pm\infty$. Suppose moreover that $f_0(v) \ge 0$. Since the (smooth) initial value is non negative, for $t \ge 0$, every point $v_m(t)$ in which  $f(v_m(t),t)=0$ is either a local minimum, and
 \be\label{mini}
 \left. \frac{\partial }{\partial v}f(v,t)\right|_{v=v_m(t)} =0, \quad \left.\frac{\partial^2 }
{\partial v^2}f(v,t)\right|_{v=v_m(t)} > 0.
 \ee
or a stationary point, and in this case
 \be\label{statio}
  \left. \frac{\partial }{\partial v}f(v,t)\right|_{v=v_m(t)} =0, \quad \left.\frac{\partial^2 }
{\partial v^2}f(v,t)\right|_{v=v_m(t)} = 0.
 \ee
Computing derivatives, the Fokker--Planck equation \fer{FP2c} can be written in the form 
 \be\label{est}
\dt f(v,t) = \frac\sigma 2 v^2 \frac{\partial^2 }
{\partial v^2}f(v,t) + \left[(2\sigma +\lambda)v - \lambda\right] \frac{\partial }{\partial v}f(v,t) +(\lambda +\sigma) f(v,t).
 \ee
Hence, evaluating \fer{est} at the point $v= v_m(t)$, and using \fer{mini} shows that, if $v_m(t)\not= 0$ is a local minimum 
 \[
\dt f(v,t)\left. \right|_{v=v_m(t)} = \frac\sigma 2 v_m(t)^2 \frac{\partial^2 }
{\partial v^2}f(v,t)\left. \right|_{v=v_m(t)}  > 0.
 \]
 This entails that the function $f(v,t)$ is increasing in time at the point $v=v_m(t)$, unless $v_m(t) = 0$. Indeed, if the local minimum is attained at $v_m(t)=0$
\[
\dt f(v,t)\left. \right|_{v=0} = 0,
 \]
and $f(0,t)$ remains equal to zero at any subsequent time.

 If now $v_m(t)$ is a stationary point, so that \fer{statio} holds, 
 \[
\dt f(v,t)\left. \right|_{v=v_m(t)} =  0,
 \]
and $f(v,t)$ remains equal to zero. Therefore
 \be\label{min3}
\min_{x \in \R} f(v,t) \ge 0,
 \ee
 and positivity follows. The proof for initial data satisfying the conditions of Theorem \ref{exi} then follows first considering a suitable smoothing of the initial data, and then using the fact that at any subsequent time $t>0$, the solution corresponding to the smoothed initial data converges to the solution of the original data when eliminating the initial smoothing. 
\medskip
\

\begin{rem}\label{v=0}
A further consequence of this analysis is that, if the initial datum vanishes on the half-line $v\le 0$, in reason of the properties of the solution at the point $v=0$, the solution at any subsequent time $t \ge 0$ will remain equal to zero on the half-line $v\le0$. 
\end{rem}

\medskip
Further results in this direction follows by studying the evolution of the mass located in the negative part of the real axis. To start with, consider that by evaluating \fer{wFP} with test functions $\phi(v) = 1, v$ one obtains that, if the initial value $f_0(v)$ vanishes for $v= \pm\infty$, the solution  to \fer{FP2c} satisfies
 \[
\frac{d}{dt} \inr  f (v,t) dv= 0, \quad  \frac{d}{dt} \inr v f (v,t) dv= \lambda \left( -\inr v f (v,t)\,dv +\inr  f (v,t) \,dv \right).
 \]
Therefore, if the (nonnegative)  initial value  of the Fokker--Planck equation  \fer{FP2c} is a  density function satisfying the normalization conditions
 \be\label{n-c}
 \inr  f_0(v)(v)\, dv=  1; \quad \inr v f_0(v)(v)\, dv=  1
 \ee
the solution $f(v,t)$ to \fer{FP2c} still satisfies conditions \fer{n-c}. In other words, if the initial datum is a probability density with unit mean, then the solution at any subsequent time remains a probability density with unit mean.

A further interesting property of the solution can be extracted by analyzing the behaviour of the mass and the mean value separately on the left and right half-line.
Let us denote by $ \rd(t)$ (respectively $\rs(t)$) the fraction of the mass distributed on the positive half-line (respectively on the negative half-line) at time $t \ge 0$, that is
 \be\label{m+-}
 \rd(t) = \id f(v,t)\, dv ;  \quad \rs(t)=\is f(v,t)\, dv.
 \ee
Let the initial value $f_0(v)\in C(\R)$ satisfy conditions \fer{n-c}. Let $H_n(v)$ be a smooth approximation to the Heaviside step function, for example the logistic function
 \[
H_n(v) = \frac 1{1 + e^{-2nv}}.
 \]
Then, equation \fer{wFP} implies, for any $t >0$
 \[
 \inr H_n(v)f (v,t)\, dv =  \inr H_n(v)f_0 (v)\, dv + 
 \]
 \[
+ \int_0^t\inr \left[\frac \sigma2  v^2 H_n''(v) -\lambda (v-1)H_n'(v)\right] \, f(v,s) \, dv\, ds.
 \]
Letting $n\to +\infty$, and considering that $H_n'(v)$ converges to a Dirac delta in zero, while $v^2H_n''(v)$ is a uniformly bounded function that converges pointwise to zero,  we obtain
 \[
 \lim_{n \to +\infty} \int_0^t\inr \left[\frac \sigma2  v^2 H_n''(v) -\lambda (v-1)H_n'(v)\right] \, f(v,s) \, dv\, ds = \int_0^t f(0,s) \, ds. 
 \]
Therefore, since for ant $t \ge 0$
 \[
\lim_{n \to +\infty}\inr H_n(v)f (v,t)\, dv = \int_0^{+\infty} f(v,t)\, dv = \rd(t),
 \]
 it follows that 
 \be\label{m+}
 \rd(t) = \rd(0) + \int_0^t f(0,s) \, ds,
 \ee
namely that the mass in the positive half-line can not decrease if the mean value is positive. 

With similar arguments it is possible to analyze the time behaviour of the parts of the mean value located on the positive and negative parts of the real line. Let us indicate these parts by $\pp(t)$ and $\pn(t)$, where 
 \be\label{me+-}
 \pp(t) = \id v f(v,t)\, dv, \quad  \pn(t) = \is vf(v,t)\, dv
 \ee
A direct computation shows that, for each time $t>0$ 
\begin{equations}\label{ide4}
&\pp(t) = \pp(0) + \int_0^t\left(-\rs(s)\pp(s) +\rd(s) \pn(s)\right) \, ds,\\
&\pn(t) = \pn(0) + \int_0^t\left((\rs(s)\pp(s) -\rd(s) \pn(s)\right)\, ds.
\end{equations}

The choice of mean value $m=1 >0$ implies $\pp(t)=|\pn|(t)+1$. Therefore using this equality into the second equation in \fer{ide4} we obtain
 \[
 \frac d{dt} |\pn(t)| = - \left((\rs(t)\pp(t) -\rd(t) \pn(t)\right)\, = -  \left(|\pn(t)| + \rs(t)|\pn(t)|\right)\,  \le  - |\pn(t)|\, .
 \]
Consequently, by Gronwall inequality we conclude that 
 \be\label{meandec}
 |\pn(t)| \le |\pn(0)| e^{-t},
 \ee
and the negative part of the mean value decays exponentially fast towards zero. We can group the previous results into the following
\medskip
\

\begin{thm}\label{properties}
Let $f_0(v)$ be a probability density in $\R$, satisfying the normalization conditions \fer{n-c}. Then, the solution $f(v,t)$ to the Fokker--Planck equation \fer{FP2c} remains a probability density for each subsequent time $t \ge0$, and satisfies conditions \fer{n-c}. Moreover, the mass $\rho_+(t)$  located on the positive part of the real line is non decreasing in time and \fer{m+} holds. Also, the part of the mean value $m_-(t)$ located on the negative part of the real axis is exponentially decreasing in time, and \fer{meandec} holds.
\end{thm}

\medskip
\

In the economic context, the consequences of Theorem \ref{properties} appear relevant.
\medskip
\
 
\begin{rem} Equation \fer{m+}, coupled with the property of mass conservation, implies that, in the particular case in which the initial data is a smooth probability density which takes values only  in the region $v\geq0$, since the mass in this region can only increase, the solution at any subsequent time $t >0$ remains a smooth probability density distributed on the same region $v\geq0$. This independently of any boundary condition one can introduce to justify mass and momentum conservation \cite{FPTT16, TT1}. This property can be easily relaxed to general probability measures initially taking values on the set $v \ge 0$. In other words, the lack of diffusion at the point $v=0$, as outlined in Remark \ref{v=0},  is enough to maintain the whole mass, initially located on the positive part of the real line, on the same set. 
\end{rem}
\medskip

\

\begin{rem}
The previous results about the time evolution of the mass  and mean value located on the set $v \ge0$ show that the part of  mass that is initially distributed on the negative half-space (the debts) moves to the region $v\geq0$, and this process is exponentially rapid in terms of the negative part of the mean value. However, since the regularity results of Theorem \ref{reg} are not uniform with respect to time, it could happen that there is accumulation of the negative fraction of the mass at the point $v=0$,  with the eventual formation of a Dirac delta in $v=0$, namely the point in which there is no diffusion.
\end{rem}
\medskip

\subsection{The stationary state}
Let us consider the
 $\Gamma$-like distribution \fer{equi2}, continuously extended to zero for $v<0$
 \be\label{equilibrio}
f_\infty(v) = 
\frac{(\mu-1)^\mu}{\Gamma(\mu)}\frac{\exp\left(-\frac{\mu-1}{v}\right)}{v^{1+\mu}} \quad {\rm if} \,\,\, v \ge 0; \qquad 
f_\infty(v) = 0  \quad {\rm if} \,\,\, v < 0.
 \ee 
  where
  \be\label{mu}
  \mu = 1 + 2 \frac{\lambda}{\sigma} >1.
 \ee
It can be easily verified that the equilibrium distribution \fer{equilibrio} achieves its  maximum value 
 \be\label{max-eq}
 \bar f_\infty = \frac{(\mu+1)^{\mu +1}}{\Gamma(\mu)(\mu-1)}\exp \left(-(\mu +1)\right)
 \ee
at the  point
\be\label{maxx}
  \bar v = \frac{\mu -1}{\mu +1}.
 \ee
 Therefore it is increasing in the interval $(0, \bar v)$ and decreasing on $(\bar v, +\infty)$.  
 Note that the value $1+ \mu$ defines the rate of decay at infinity of the power tailed distribution \fer{equilibrio}. Consequently
 \be\label{mmo}
 \int_\R |v|^r f_\infty(v) \, dv < \infty
 \ee
if and only if $r < \mu$.

 Then, owing to elementary properties of the Gamma function, it is immediate to conclude that, provided $\mu >2$, the second moment of the steady state is bounded, and
 \be\label{mom-e}
 \int_\R f_\infty(v)\, dv = 1; \quad \int_\R v\,f_\infty(v)\, dv = 1; \quad \int_\R v^2 f_\infty(v)\, dv = \frac{\mu-1}{\mu -2}.
 \ee
It follows that, if the initial value for the Fokker--Planck equation \fer{FP2c} posed in the whole space $\R$ is a probability density function of mean value equal to one,  $f_\infty(v)$ is a smooth probability density with the same mean value, which in addition satisfies the Fokker--Planck equation \fer{FP2c} on $\R$. If in addition $\mu >2$, and the initial value has the second moment bounded, then the second moment of the solution converges exponentially towards the second moment of $f_\infty$. 

 For $n \in \N_+$ let us define
 \[
 M_n(t)= \int_{\R_+} v^n f (v,t) dv.
 \]
Then \cite{TT1}
 \be\label{2-th}
\frac{d}{dt} M_2(t)= (\sigma - 2\lambda) M_2(t) + 2\lambda.
  \ee
Hence, the value of the second moment stays bounded when $\sigma < 2\lambda$ (or, what is the same $\mu >2$), while it diverges in the opposite case.  In the former case, solving equation \fer{2-th}
 we obtain
\be\label{m2}
M_2(t)= e^{(\sigma-2\lambda)t}\left(M_2(0) +\frac{2\lambda }{\sigma-2\lambda} \right) +\frac{2\lambda}{2\lambda -\sigma} ,
\ee
 which implies
 \[
 \lim_{t \to \infty} M_2(t) = \frac{2\lambda}{2\lambda -\sigma}.
 \]
 Thus, $f_\infty(v)$ is the (unique) steady state of the Fokker--Planck equation with moments satisfying \fer{mom-e}. 
 This clearly indicates that one could expect that, even starting with a probability density defined on the whole $\R$, but with positive mean value (equal to one in our case), the solution to the initial value problem will converge in time towards the equilibrium \fer{equilibrio}. A rigorous proof of this property will be presented in the next section.
 \begin{rem}
It is clear that  the evolution of the principal moments of the solution to the Fokker--Planck equation \fer{FP2c} can be obtained recursively, and explicitly evaluated at the price of an increasing length of computations. 
\end{rem}

\section{Convergence to equilibrium}\label{conv}

\subsection{Fourier based metrics}
As shown in Section \ref{Bol}, in reason of the positivity property, and mass and momentum conservation of the solution of the Fokker--Planck equation \fer{FP2c}, one can always assume that both the solution and the steady state are probability densities satisfying \fer{n-c}. This remark allows to use metrics for probability distributions to study convergence to equilibrium. This is a method that in kinetic theory of rarefied gases goes back to  \cite{GTW}, where convergence to equilibrium for the Boltzmann equation for Maxwell pseudo-molecules was studied in terms of a metric for Fourier transforms (cf. also \cite{Carrillo:2007, MaTo07, TV} for further applications). 

For a given constant $s >0$ let $\mathcal M_{s}$ be the set of probability measures $\mu$ on the Borel subsets of $\R$ such that
 \[
 \int_\R |v|^s\, \mu(dv) < \infty,
 \]
and let $\mathcal F_s$ be the set of Fourier transforms of probability distributions $\mu$ in $M_s$. 
A useful metric in $\mathcal F_s$ has been introduced in \cite{GTW} in connection with the Boltzmann equation for Maxwell molecules, and subsequently applied in various contexts, which include kinetic models for wealth distribution \cite{PT13}.  For a given pair of random variables $X$ and $Y$ distributed according to $\phi$ and $\psi$ this metric reads
\be\label{me-inf}
 d_s(X,Y) = d_s(\phi,\psi) = \sup_{\xi\in\R} \frac{|\widehat \phi(\xi) - \widehat \psi(\xi)|}{|\xi|^s},
 \ee
As shown in \cite{GTW}, the metric $d_s(\phi,\psi)$ is finite any time the probability distributions $\phi$ and $\psi$ have equal moments up to $[s]$, namely the entire part of $s\in \R_+$, or equal moments up to $s-1$ if $s \in \N$, and it is equivalent to the weak$\null^*$ convergence of measures for all $s >0$. Among other properties, it is easy to see \cite{GTW, PT13} that, for any random variable $Z$ independent of $X$ and $Y$ and for any  constant $c$ 
 \begin{equations}\label{sca1}
 & d_s(X+Z, Y+Z)\le d_s(X,Y), \\
 & d_s(c X, c Y) =  |c|^s d_s(X,Y).
 \end{equations}
These properties classify $d_s$ as an ideal probability metric in the sense of Zolotarev \cite{Zolo}. 

Few years after the publication of \cite{GTW}, Baringhaus and Gr\"ubel \cite{BG}, in connection with the study of convex combinations of random variables with random coefficients, considered a Fourier metric similar to \fer{me-inf}, defined by
 \be\label{me-1}
 D_s(X,Y) = D_s(\phi,\psi) = \int_\R \frac{|\widehat \phi(\xi) - \widehat \psi(\xi)|}{|\xi|^{1+s}}\, d\xi.
 \ee
As shown in \cite{BG}, also  $D_s$ as an ideal probability metric in the sense of Zolotarev, and for $1<s<2$ the space $\tilde{\mathcal{F}_s}\subset \mathcal F$  of probability distributions satisfying \fer{n-c} endowed with the metric $D_s$ is complete. 

It can be verified that the metrics $d_s$ and $D_s$ are strictly connected. In particular, if $r < s$, $D_r(\phi,\psi) \le c(r,s) d_s(\phi,\psi)^{r/s}$, where $c(r,s)$ is a positive constant which depends only on $r,s$. 

Indeed, since $|\widehat \phi(\xi)| \le 1$, $| \widehat \psi(\xi)| \le 1$, for any given positive constant $R$
 \[
 \int_{|\xi| >R} \frac{|\widehat \phi(\xi) - \widehat \psi(\xi)|}{|\xi|^{1+r}}\, d\xi \le  \int_{|\xi| >R}\frac{2}{|\xi|^{1+r}}\, d\xi = \frac 4{rR^r}.
 \]
On the  other hand, on the interval $|\xi \le R|$, for $s>r$ it holds
 \[
 \int_{|\xi| \le R} \frac{|\widehat \phi(\xi) - \widehat \psi(\xi)|}{|\xi|^{1+r}}\, d\xi =  \int_{|\xi| \le R} \frac{|\widehat \phi(\xi) - \widehat \psi(\xi)|}{|\xi|^{s}}\cdot \frac{1}{|\xi|^{1 + r -s}}\, d\xi \le
 \]
 \[
  d_s(\phi, \psi) \int_{|\xi| \le R}\frac{1}{|\xi|^{1 + r -s}}\, d\xi =  2 \,d_s(\phi, \psi)\frac{R^{s-r}}{s-r}. 
 \]
Therefore, for for any given positive constant $R$
 \[
 D_r(\phi,\psi) \le 2 \,d_s(\phi, \psi)\frac{R^{s-r}}{s-r} +  \frac 4{rR^r},
 \]
and, optimizing over $R$ we obtain, for $s>r$
 \be\label{est5}
 D_r(\phi,\psi) \le c(r,s) d_s(\phi,\psi)^{r/s},
 \ee
where
 \be\label{con5}
 c(r,s) = 2^{2-r/s} \frac{s}{r(s-r)}.
 \ee 
This allows to conclude that for $1<r<2$, and for $s>r$, the space $\tilde{\mathcal{F}_s}$ endowed with the metric $d_s$ is complete.

New metrics on $\mathcal F_s$ can be introduced according to the following definition. Let $p \ge 1$, and $s >0$. For a given pair of random variables $X$ and $Y$ distributed according to $\phi$ and $\psi$ we define
\be\label{Ds}
D_{s,p}\,(X,Y) = D_{s,p}\,(\phi,\psi)=\left[\int_\R |\xi|^{-(ps+1)}|\widehat\phi(\xi)-\widehat\psi(\xi)|^p d \xi\right]^\frac{1}{p}.
\ee
The metric $D_s$ corresponds to $D_{s,1}$, while the metric $d_s$ is obtained by taking the limit $p\to \infty$ of $D_{s,p}$. Moreover, for any given value of the constant $p$, the $D_{s,p}$ metric is an ideal probability metric in the sense of Zolotarev. Proceeding as before, it is immediate to show that these metrics satisfy an inequality similar to \fer{est5}
\be\label{est6}
 D_r(\phi,\psi) \le c(p,r,s) d_s(\phi,\psi)^{r/s},
 \ee
where
 \be\label{con6}
 c(p,r,s)= 2^{1-r/s}\left[\frac{2s}{pr(s-r)}\right]^{\frac{1}{p}}.
 \ee 
In addition, it can be shown that the $D_{s,p}$-metrics are strictly related each other. 
In fact, if $p <q$ and $r<s$, by similar methods one proves that there exists a finite explicitly computable constant such that the following estimate holds
\be\label{est7}
 D_{r,p}(\phi,\psi) \le c(p,q,r,s) D_{s,q}(\phi,\psi)^{r/s}.
 \ee

A distinguished case is obtained by fixing $p=2$. Then the $D_{s,2}$ metric
\be\label{D2}
D_{s,2}\,(\phi,\psi)=\left[\int_\R |\xi|^{-(2s+1)}|\widehat\phi(\xi)-\widehat\psi(\xi)|^2 d \xi\right]^\frac{1}{2}.
\ee
coincides with the distance between $\phi$ and $\psi$ in the homogeneous Sobolev space of fractional order with negative index $\dot H_{-q}$, with $q = s + 1/2$, where, for $h \in \dot H_{-q}$
 \be\label{h2}
  \| h\|_{\dot H_{-q}} = \int_\R |\xi|^{-2q}|\widehat h(\xi)|^2 \, d\xi.
 \ee
 
\subsection{Convergence in Fourier metric}\label{Fou} 
Convergence to equilibrium of the solution to the Fokker--planck equation \fer{FP2c} in the metric $D_{s,2}$ is an easy consequence of the result of Theorem \ref{reg}. Indeed, looking at its proof it is immediate to notice that all computations leading to formula \fer{gro-t} still holds when $r<0$. Moreover, thanks to the linearity of the Fokker--Planck equation \fer{FP2c}, formula \fer{gro-t} remains valid if we substitute $f(v,t)$ with the difference $f(v,t) -f_\infty(v)$. Hence, by setting $r = -(s+1/2)$ we obtain
 \begin{equations} \label{deca}
 & D_{s,2}\,(f(t),f_\infty) = \left[\int_\R |\xi|^{-(2s+1)}|\widehat f(\xi)-\widehat f_\infty(\xi)|^2 d \xi\right]^\frac{1}{2} \le \\
 & \exp\left\{ -\frac s2\left( (1-s)\sigma) + 2\lambda\right) \right\}\left[\int_\R |\xi|^{-(2s+1)}|\widehat f_0(\xi)-\widehat f_\infty(\xi)|^2 d \xi\right]^\frac{1}{2} =\\
& \exp\left\{ -\frac s2\left( (1-s)\sigma) + 2\lambda\right) \right\}  D_{s,2}\,(f_0,f_\infty).
\end{equations}
Therefore, if the exponent is negative, there is exponential convergence in $D_{s,2}$-metric of the solution $f(v,t)$ towards the steady distribution  $f_\infty(v)$. This happens if
 \be\label{con44}
 s < 1 + 2\frac{\lambda}\sigma = \mu
 \ee
where the constant $\mu$ has been defined in \fer{mu}, and characterizes the decay at infinity of the stationary distribution $f_\infty(v)$. Note that, since $\mu >1$, by taking $s=1$ we obtain that convergence to equilibrium holds for all initial values satisfying \fer{n-c} at a rate $2\lambda$, which results to be independent of the coefficient of diffusion $\sigma$ of the Fokker--Planck equation \fer{FP2c}. Hence we have

\begin{thm}\label{conv-exp}
Let $f_0(v)$ be a probability density in $\R$ satisfying \fer{n-c}, and such that $D_{s,2}(f_0,f_\infty)$ is finite for some $s <\mu$, where $\mu$ is defined in \fer{mu}. Then, the solution to the Fokker--Planck equation \fer{FP2c} posed in the whole space $\R$ is exponentially converging to the equilibrium density $f_\infty$ in $D_{2,s}$-metric and the following decay holds
 \be\label{decay-t}
D_{2,s}(f(t),f_\infty) \leq  \exp\left\{ -\frac s2\left( (1-s)\sigma) + 2\lambda\right) \right\} D_{2,s}(f_0,f_\infty).
\ee
\end{thm}

It is immediate to verify that the rate of decay to equilibrium is maximum when $s = \mu/2$. In this case, provided $D_{2,\mu/2}(f_0,f_\infty)<\infty$
 \be\label{decay-mu}
D_{2,\mu/2}(f(t),f_\infty) \leq  \exp\left\{ -\frac {\sigma\mu^2}8 \right\} D_{2,\mu/2}(f_0,f_\infty).
\ee

\subsection{The monotonicity of relative entropy} \label{ent-mon}
The result of  Section \ref{Fou} shows that, at least in a weak sense, there is exponential convergence of the solution to the Fokker--Planck equation \fer{FP2c} posed in the whole space towards the unique steady state $f_\infty(v)$ defined in \fer{equilibrio}.  
 As usual for this type of equations \cite{FPTT16}, to obtain stronger convergence results, tipically in $L^1(\R)$, a classical method is to resort to the study of the time decay of various Lyapunov functionals involving  the solution $f(v,t)$ and the steady state.

Among these functionals, a leading rule is usually assumed by the relative Shannon entropy $H(f,h)$,  where, for any given pair of probability densities $f,h$
\be\label{relH}
 H(f,h)= \int_{\R} f(v) \log \frac {f(v) }{h(v) }\, dv.
 \ee
However, since in our case the steady state \fer{equilibrio} is supported on the positive real line, while the initial value is in general supported on the whole real line, the Shannon entropy $H(f(t), f_\infty)$  of the solution $f(v,t)$ to the Fokker--Planck equation \fer{FP2c} relative to the steady state $f_\infty(v)$ is unbounded, and consequently useless.

A related relative entropy which appears more appropriate to treat the present problem is the so-called Jensen-Shannon entropy, introduced by Lin in \cite{lin}. Given the pair of probability densities $f,h$, and a constant $0< \alpha < 1$, the Jensen-Shannon entropy $H_\alpha$ of $f$ relative to $h$ is defined by 
\be\label{Je-Sh}
H_\alpha(f,h)=\int_{\mathbb{R}}f(v) \log \frac{f(v)}{\alpha f(v) + (1-\alpha)h(v)}\, dv.
\ee
Note that, since the convex combination $\alpha f +(1-\alpha)h$ of the two probability densities $f$ and $h$ is still a probability density, say $h_\alpha$, the Jensen-Shannon entropy $H_\alpha$ of $f$ relative to $h$ is simply the Shannon entropy of $f$ relative to $h_\alpha$. The main properties of these entropies have been studied in \cite{lin} (cf. also \cite{FPTT16}). In particular, thanks to Lemma $\rm 27$ in \cite{FPTT16} the Jensen--Shannon entropy of two probability densities is always bounded.

Let  $g(v,t) = \alpha f(v,t) + (1-\alpha) f_\infty(v)$. Thanks to \fer{n-c} and \fer{mom-e}, it follows that $g(v,t)$ is a probability density of unit mean. Moreover, since both $f(v,t)$ and $f_\infty(v)$ are solutions to the linear Fokker--Planck equation \fer{FP2c}, $g(v,t)$ is itself a solution to \fer{FP2c}.
Note that for $v\leq0$, $f(v)/g(v) = 1/{\alpha} $. Moreover, if $v>0$,  $f(v)/g(v)\leq 1/\alpha$. Since for $ r \ge 0$ the function $r\log r$ is bounded from below,  writing  
$$
\Ha=\int_{\R}\left( \frac{f(v)}{g(v)} \log \frac{f(v)}{g(v)}\right) g(v)\, dv,
$$
it is immediate to conclude that $\Ha$ is well defined and bounded from above and below independently of the regularity of the initial data.

In what follows, to avoid inessential difficulties in the forthcoming computations, we will assume that the initial density $f_0(v)$ (and consequently the solution $f(v,t)$), is smooth and has enough moments bounded.  

To compute the the evolution of the Jensen--Shannon entropy, let us first remark that
\[
\int_{-\infty}^0 f(v,t) \log \frac{f(v,t)}{\alpha f(v,t) + (1-\alpha)f_\infty(v)}\, dv = \log\frac 1\alpha \int_{-\infty}^0 f(v,t)\, dv .
\]
Since, according to \fer{m+} the mass in the negative half-line can not increase, and $\log\frac 1\alpha >0$, the part of Jensen--Shannon entropy relative to the domain $v \le 0$ is nonincreasing in time. 

On the set $(\var, +\infty)$ we have
\begin{equations}\label{J-S}
\frac{d}{dt}\int_\var^\infty f\log\frac fg\, dv =&\int_\var^\infty \bigg\{
 \log  \frac{f}{g}\dvv (v^2 f)- \frac{f}{g}\dvv (v^2 g)\bigg\}dv\\
 &+\int_\var^\infty \bigg\{ \log  \frac{f}{g}\dv [(v-1) f]- \frac{f}{g}\dv [(v-1) g]  \bigg\}dv.
 \end{equations}
Using the identity $f/g= (v^2f)/(v^2g)$, that clearly holds when $v \in (\var, +\infty)$, integration by parts  gives (cf. the proof of Proposition $\rm 25$ in \cite{FPTT16})
\begin{equations}
&\int_{\var}^{+\infty}\bigg\{\log  \frac{v^2f}{v^2g}\dvv (v^2 f)-\frac{v^2f}{v^2g}\dvv (v^2 g)\bigg\}\, dv
=\\
& \bigg[\log  \frac{f}{g}\dv (v^2f)- \frac{f}{g}\dv(v^2 g)\bigg]_{\var}^{+\infty} -\int_{\var}^{+\infty}v^2{f}\left[\dv \log\frac{f}{g}\right]^2\, dv.
\end{equations}
The contribution of the border term at infinity can be easily shown to vanish provided $f(v,t)$ has moments bounded of order $2+\delta$ for some $\delta >0$. Indeed, given $p, q$ conjugate exponents, namely such that $1/p+1/q =1$
 \[
 \left| \log  \frac{f}{g}\dv (v^2f)\right| =  \left|\left(\frac{f}{g}\right)^{1/q}\log  \frac{f}{g} \left(\frac{v^2g}{v^2f}\right)^{1/q}\dv (v^2f)\right| \le
C_q (v^2g)^{1/q}p \left|\frac{\partial (v^2f)^{1/p}}{\partial v}\right|.
 \]
In the previous inequality we defined 
 \[
 C_q = \sup \left|\left(\frac{f}{g}\right)^{1/q}\log  \frac{f}{g}\right|,
 \]
which is bounded in reason of the fact that $f/g \le 1/\alpha$. 
Moreover, by H\"older inequality, whenever $p/q \le \delta/2$
 \[
 \int_\R (v^2f)^{1/p}\, dv = \int_\R (v^2f)^{1/p}(1+v^2)^{1/q}(1+v^2)^{-1/q}  \, dv  \le
 \]
 \[
  \left( \int_\R v^2(1+v^2)^{p/q}f\, dv \right)^{1/p}\left( \int_\R (1+v^2)^{-1}\, dv \right)^{1/q} \le C.
 \]
Consequently, as soon as  $p/q \le \delta/2$, both the smooth functions $v^2g$ and $(v^2f)^{1/p}$ are integrable, and
 \be\label{limo}
 \lim_{v \to \infty} (v^2g)^{1/q}\left|\frac{\partial (v^2f)^{1/p}}{\partial v}\right| = 0.
 \ee
Analogous arguments can be used to prove that 
 \be\label{limo2}
  \lim_{v \to \infty} \frac{f}{g}\dv(v^2 g)=0.
 \ee
On the other extremal point, the choice $p=q=2$  gives
 \[
  \left| \log  \frac{f}{g}\dv (v^2f)\right| \le 2 C_2 (v^2g)^{1/2} \left|\frac{\partial( v f^{1/2})}{\partial v}\right| = 2 C_2 \left|vg^{1/2}\left(v \frac{\partial  f^{1/2}}{\partial v} + f^{1/2}\right)\right|.
 \]
Then, considering that both $f$ and $g$ are smooth, and $f^{1/2} \in L^2(\R)$, one obtains
 \be\label{lim}
 \lim_{v \to 0} (v^2g)^{1/2}\left|\frac{\partial (v^2f)^{1/2}}{\partial v}\right| = 0,
 \ee
and
\be\label{lim2}
  \lim_{v \to 0} \frac{f}{g}\dv(v^2 g)=0.
 \ee 
Let us consider now the second integral into \fer{J-S}. Integrating by parts first on the interval $(\var, 1-\var)$, and using the identity $f/g= (v-1)f/[(v-1)g]$ we obtain
 \begin{equations}
&\int_{\var}^{1-\var} \bigg( \log  \frac{f}{g}\dv [(v-1) f]- \frac{f}{g}\dv [(v-1) g]  \bigg)\, dv = \\
& \bigg[ \log  \frac{f}{g} (v-1) f- (v-1) f  \bigg]_{\var}^{1-\var} = \bigg[ \frac{f}{g} \log  \frac{f}{g} (v-1) g - (v-1) f . \bigg]_{\var}^{1-\var}  \end{equations}
Hence, since the quantity $(f/g)\log(f/g)$ is uniformly bounded from above and below
 \[
  \lim_{v \to 1} \left(\frac{f}{g} \log  \frac{f}{g} (v-1) g - (v-1) f \right) =0.
  \]
Moreover, since 
\[
 \lim_{v \to 0} \frac{f(v)}{g(v)} = \frac 1\alpha,
 \]
we obtain
 \[
\lim_{v \to 0}\left( \log  \frac{f(v)}{g(v)} (v-1) f(v) - (v-1) f (v) \right) = f(0)\left( 1 + \log \alpha \right). 
 \]
This implies
\[
\int_0^1 \bigg( \log  \frac{f}{g}\dv [(v-1) f]- \frac{f}{g}\dv [(v-1) g]  \bigg)\, dv = - f(0)\left( 1 + \log \alpha \right)\]
Similar computations then give
\[
\int_1^{+\infty} \bigg( \log  \frac{f}{g}\dv [(v-1) f]- \frac{f}{g}\dv [(v-1) g]  \bigg)\, dv = 0.
\]
Grouping the various pieces, we conclude that the Jensen--Shannon entropy $H_\alpha(f(t),f_\infty)$ is nonincreasing in time.  We proved

\medskip

\begin{thm}\label{decay-JS}
Let $f_0(v)$ be a smooth probability density in $\R$ satisfying \fer{n-c}, and such that its moments up to $2+\delta$ are finite for some $\delta >0$. Then, for any $0<\alpha<1$, the Jensen--Shannon entropy $H_\alpha(f(t), f_\infty)$ of the solution to the Fokker--Planck equation \fer{FP2c} relative to the equilibrium solution is monotonically nonincreasing, and the following decay holds
\be\label{dec}
H_\alpha(f(t),f_\infty) = H_\alpha(f_0,f_\infty)  - \int_0^t f(0,s) \, ds -  \int_0^t  \int_{0}^{+\infty} v^2 f(v,s)\left[ \dv \log\frac{f(v,s)}{g(v,s)} \right]^2\, dv\, ds.
 \ee
\end{thm}

\subsection{The monotonicity of Hellinger distance}\label{hel-mon} 
A second interesting functional that has been shown to be monotonically decreasing along the solution to Fokker--Planck type equations \cite{FPTT16} is the Hellinger distance. For any given pair of probability densities $f$ and $h$  defined on  $\R$, the Hellinger distance $d_H(f,h)$ is \cite{Zolo} 
\be\label{hel1}
d_{H} (f,h)= \left(\int_\R \left(\sqrt {f(v) } -\sqrt {h(v) }\right )^2\ d v\right )^{\frac 12}.
\ee 
In what follows, in analogy with the definition of  Jensen-Shannon entropy, defined in \fer{Je-Sh}, we will define, for $0<\alpha<1$ the $\alpha$-Hellinger distance of $f$ and $h$ by
\be\label{a-hel}
d_{H, \alpha} (f,h)^2 =\int_{\R}\left(\sqrt {f(v) } -\sqrt{\alpha f(v) + (1-\alpha)h(v)}\right)^2\, dv,
\ee
and we will study  the time-evolution of the square of the $\alpha$-Hellinger distance between the solution $f(v,t)$ of the Fokker--Planck equation \fer{FP2c}, and the equilibrium density $f_\infty(v)$, namely the square of the Hellinger distance between $f(v,t)$ and  $g(v,t) = \alpha f(v,t) + (1-\alpha) f_\infty(v)$. 

As in Section \ref{ent-mon}, we will assume that the initial density $f_0(v)$ (and consequently the solution $f(v,t)$), is smooth and has enough moments bounded. Moreover, since most of the computations that follow are analogous to the computations of Section \ref{ent-mon}, we will only outline the differences. 

To compute the the evolution of the square of the $\alpha$-Hellinger distance, let us first remark that
\[
\int_{-\infty}^0 \left(\sqrt {f(v) } -\sqrt{\alpha f(v) + (1-\alpha)f_\infty(v)}\right)^2\, dv = \left(1 -\sqrt\alpha\right)^2 \int_{-\infty}^0 f(v,t)\, dv .
\]
Therefore, since according to \fer{m+} the mass in the negative half-line can not increase, and $\left(1 -\sqrt\alpha\right)^2>0$, the part of the square of the $\alpha$-Hellinger distance relative to the domain $v \le 0$ is nonincreasing in time. 

On the set $(\var, +\infty)$ we have
\begin{equations}\label{H1}
& \frac{d}{dt}\int_\var^\infty (\sqrt f - \sqrt g)^2 \, dv = \\
&\int_\var^\infty \bigg\{ \left(1 - \sqrt{\frac gf}\right)\dvv (v^2 f) +  \left(1 - \sqrt{\frac fg }\right)\dvv (v^2 g) 
 \bigg\}\, dv\\
 &+\int_\var^\infty \bigg\{  \left(1 - \sqrt{\frac gf}\right)\dv [(v-1) f]+  \left(1 - \sqrt{\frac fg}\right) \dv [(v-1) g]  \bigg\}\, dv.
 \end{equations}
Using the identity $f/g= (v^2f)/(v^2g)$, that clearly holds when $v \in (\var, +\infty)$, integration by parts  gives (cf. the proof of Proposition $\rm 25$ in \cite{FPTT16})
\begin{equations}\label{H2}
&\int_{\var}^{+\infty}\bigg\{ \left(1 - \sqrt{\frac{v^2g}{v^2f}}\right)\dvv (v^2 f) +  \left(1 - \sqrt{\frac{v^2 f}{v^2 g}}\right)\dvv (v^2 g) 
 \bigg\} \, dv
=\\
& \bigg[\left(1 - \sqrt{\frac{v^2g}{v^2f}}\right) \dv (v^2f)+ \left(1 - \sqrt{\frac{v^2 f}{v^2 g}}\right)\dv(v^2 g)\bigg]_{\var}^{+\infty} -\frac 12 \int_{\var}^{+\infty}v^2\sqrt{fg}\left[\dv \log{ \frac{f}{g}}\right]^2\, dv.
\end{equations}
Proceeding as in the proof of monotonicity of Jensen-Shannon entropy, the contribution of the border term at infinity in \fer{H2} can be easily shown to vanish provided $f(v,t)$  possesses moments bounded of order $3+\delta$ for some $\delta >0$. Indeed, it is enough to follow the proof of Section \ref{ent-mon} by choosing $p=q=2$. 

 On the other extremal point,  considering that both $f$ and $g$ are smooth, and $f^{1/2} \in L^2(\R)$, one obtains
 \be\label{limH}
 \lim_{v \to 0} (v^2g)^{1/2}\left|\frac{\partial (v^2f)^{1/2}}{\partial v}\right| = 0,
 \ee
and, since $f/g$ is bounded,
\be\label{limH2}
  \lim_{v \to 0} \sqrt{\frac{f}{g}}\dv(v^2 g)=0.
 \ee
Let us consider now the second integral into \fer{H1}. Integrating by parts first on the interval $(\var, 1-\var)$, and using the identity $f/g= (v-1)f/[(v-1)g]$ we obtain
 \begin{equations}
&\int_{\var}^{1-\var} \bigg( \left(1 - \sqrt{\frac gf}\right)\dv [(v-1) f]+  \left(1 - \sqrt{\frac fg}\right) \dv [(v-1) g]   \bigg)\, dv = \\
& \bigg[ (v-1)(f+g) - 2(v-1)\sqrt{fg}  \bigg]_{\var}^{1-\var}.
 \end{equations}
Hence
 \[
  \lim_{v \to 1} \left( (v-1)(f+g) - 2(v-1)\sqrt{fg} \right) =0,
  \]
and
 \[
\lim_{v \to 0} \left( (v-1)(f+g) - 2(v-1)\sqrt{fg} \right) = - f(0)\left( 1 -\sqrt\alpha \right)^2. 
 \]
This implies
\[
\int_0^1 \bigg( (v-1)(f+g) - 2(v-1)\sqrt{fg}  \bigg)\, dv = 
 f(0)\left( 1 -\sqrt\alpha \right)^2. 
\]
Similar computations then give
\[
\int_1^{+\infty} \bigg( (v-1)(f+g) - 2(v-1)\sqrt{fg} \bigg)\, dv = 0.
\]
Grouping the various pieces, we conclude that the square of the $\alpha$-Hellinger distance is  nonincreasing in time. We have

\medskip

\begin{thm}\label{decay-H}
Let $f_0(v)$ be a smooth probability density in $\R$ satisfying \fer{n-c}, and such that its moments up to $3+\delta$ are finite for some $\delta >0$. Then, for any $0<\alpha<1$, the $\alpha$-Hellinger distance $d_{H,\alpha}(f(t), f_\infty)$ between the solution to the Fokker--Planck equation \fer{FP2c} and the equilibrium solution is monotonically nonincreasing, and the following decay holds
\be\label{decH}
d_{H,\alpha}(f(t),f_\infty) = d_{H,\alpha}(f_0,f_\infty) - \frac 12 \int_0^t  \int_{0}^{+\infty} v^2 \sqrt{f(v,s)g(v,s)}\left[ \dv \log \frac{f(v,s)}{g(v,s)} \right]^2\, dv\, ds.
 \ee
\end{thm}

\medskip

Note that, at difference with the Jensen--Shannon entropy, the behavior of the solution at the point $v=0$ does not enter into the expression of the entropy production. 

As we shall see in the next Section, the monotonicity of Hellinger distance can coupled with the monotonicity of Jensen--Shannon entropy to obtain decay without rate of some $\alpha$-Hellinger  distance towards zero. 

\subsection{The decay of the $\alpha$-Hellinger distance}
 In general,  precise lower bounds for the entropy production of the Jensen--Shannon entropy are difficult to obtain. The main obstacle comes from the fact that, at difference with the case treated in \cite{TT1}, where the support of the initial value coincides with the support of the steady state a Chernoff-type inequality \cite{C, Kla, FPTT16}  connecting the relative entropy production \fer{dec} found in Theorem \ref{decay-JS} with the Hellinger distance \fer{a-hel} ( cf. \cite{JB, FPTT16}) is not available here. Nevertheless,  we can still resort to Chernoff inequality to obtain a convergence result in $\alpha$-Hellinger distance. Thanks to the identity
 \be\label{idea}
f \left( \dv \log\frac{f}{g} \right)^2 = f\left( \frac 1g \dv g - \frac 1{f} \dv f \right)^2 = \frac{(1-\alpha)^2}{\alpha^2} g\left( \frac 1g \dv g - \frac 1{f_\infty} \dv f_\infty \right)^2 \, \frac{f_\infty^2}{fg},
 \ee 
and to the upper bound
 \[
 f = \frac 1\alpha\cdot \alpha f \le \frac 1\alpha g, 
 \] 
the integral part of the entropy production of the Jensen--Shannon entropy can be bounded  as 
 \begin{equations}\label{3w}
 & \int_0^\infty  v^2 f(v,s)\left[ \dv \log\frac{f(v,s)}{g(v,s)} \right]^2\, dv = \\ &\frac{(1-\alpha)^2}{\alpha^2}
  \int_0^\infty  v^2 g(v,s)\left[ \dv \log\frac{g(v,s)}{f_\infty(v)} \right]^2 \frac{f_\infty(v)^2}{f(v,s)g(v,s)}\, dv \ge \\
  &\frac{(1-\alpha)^2}{\alpha}\int_0^\infty  v^2 g(v,s)\left[ \dv \log\frac{g(v,s)}{f_\infty(v)} \right]^2 \left(\frac{f_\infty(v)}{g(v,s)}\right)^2\, dv = \\
& 4\, \frac{(1-\alpha)^2}{\alpha}\int_0^\infty  v^2 f_\infty(v)\left[ \dv \sqrt{\frac{g(v,s)}{f_\infty(v)}} \right]^2\,\left(\frac{f_\infty(v)}{g(v,s)}\right)^2\, dv =\\
&4\, \frac{(1-\alpha)^2}{\alpha}\int_0^\infty  v^2 f_\infty(v)\left[ \dv \sqrt{\frac{f_\infty(v)}{g(v,s)}} \right]^2\,\, dv.
 \end{equations}
For the last equality in \fer{3w} we refer to \cite{FPTT16,JB}.

We can now apply Chernoff inequality with weight, in the form proven in \cite{FPTT16}.
\medskip
\
\begin{thm}[\cite{FPTT16}]\label{C}
Let $X$ be a random variable distributed with density $f_\infty(v)$, $v \in I \subseteq \R$, where the probability density function $f_\infty$ satisfies the differential equality
\be\label{staz-22}
\frac{\partial }{\partial v}\left(\kappa(v)  f_\infty \right) + (v -m)\,f_\infty = 0, \quad v\in I.
 \ee
If  the function $\phi$ is absolutely continuous  on $I$ and $\phi(X)$ has finite variance, then  
\be\label{chernoff-gen}
 Var[\phi(X)] \le E\left\{\kappa(X)[\phi'(X)]^2\right\}
\ee
 with equality if and only if $\phi(X)$ is linear in $X$.
\end{thm}

\medskip

We apply Theorem \ref{C} with $I = \R_+$, $\kappa(v) = \sigma/(2\lambda)v^2$, and density $ f_\infty(v)$, which is such that \fer{staz-22} holds in $\R_+$. Moreover
 \[
 \phi(v) = \sqrt{\frac{f_\infty(v)}{g(v)}}.
  \]
By \fer{chernoff-gen}
 \begin{equations}\label{bene}
 &\int_0^\infty \kappa(v) f_\infty(v) \left[ \dv  \sqrt{\frac{f_\infty(v)}{g(v,s)}} \right]^2 \, dv \ge \\
& \int_0^\infty \left[ \sqrt{\frac{f_\infty(v)}{g(v,s)}}  - \int_0^\infty \sqrt{\frac{f_\infty(w)}{g(w,s)}}\,  f_\infty(w) \,dw \right]^2  f_\infty(v)\, dv = \\
& \|h(s)\|_{L_1}\left[ 1 - \left(\int_0^\infty \sqrt{\bar h(v,s) f_\infty(v)}\right)^2 \right] .
\end{equations}
In \fer{bene} we defined
 \be\label{hh}
 h(v,s) = \frac{f_\infty(v)^2}{g(v,s)},  
 \ee
 and with $\bar h(v,s)$ the probability density on $\R_+$ given by
  \be\label{pp}
 \bar h(v,s) = \frac{h(v,s)}{\|h(s)\|_{L_1}}.
   \ee
  Note that, since by definition $ f_\infty(v)/g(v,s) \le 1/(1-\alpha)$, and by Cauchy--Schwartz inequality
   \[
 1=  \int_0^\infty f_\infty(v) \, dv =  \int_0^\infty \frac{f_\infty(v)}{\sqrt{g(v,s)}} \sqrt{g(v,s)}\, dv   \le \|h(s)\|_{L_1}^{1/2} 
 \left(\int_0^\infty g(v,s) \, dv\right)^{1/2},
 \]
 for all $s \ge 0$  it holds 
  \be\label{bbc}
  1 \le \|h(s)\|_{L_1} \le \frac 1{1-\alpha}.
     \ee
 On the other hand, as proven in \cite{JB}, for any given pair of probability densities $f,g$
  \be\label{Joh}
  1- \left(\int_0^\infty \sqrt{f(v)g(v)}\, dv \right)^2 \ge \frac 12 \,d_H(f,g)^2.
  \ee
In conclusion we obtain that on $\R_+$ the entropy production of the Jensen--Shannon entropy satisfies the lower bound
 \be\label{low-b}
 \int_0^\infty  v^2 f(v,s)\left[ \dv \log\frac{f(v,s)}{g(v,s)} \right]^2\, dv \ge \frac{\alpha \sigma}{4(
1-\alpha)^2\lambda}\, d_H(h(s),  \|h(s)\|_{L_1}f_\infty(s))^2.
 \ee
Note that in \fer{low-b} the coefficient is independent of  time. Substituting into \fer{dec},  \fer{low-b} implies that  
 \[
 \int_0^\infty d_H(h(s), \|h(s)\|_{L_1} f_\infty(s))^2\, ds \le  \frac{\alpha \sigma}{2(
1-\alpha)^2\lambda} H_\alpha (f_0,f_\infty).
 \]
Consequently,  the sequence $\{  d_H(h(t), \|h(t)\|_{L_1} f_\infty(t)) \}_{t \ge 0}$ contains a subsequence 

\noindent
$\{  d_H( h(t_n),  \|h(t_n)\|_{L_1}f_\infty(t_n)) \}_{n\ge 0}$ such that, as $n \to \infty$,  $t_n \to \infty$, and 
 \be\label{limit6}
 \lim_{n \to \infty} d_H( h(t_n), \|h(t_n)\|_{L_1} f_\infty) = 0. 
 \ee
 Now, consider that, for any given nonnegative $L_1$-functions $p(v)$ and $q(v)$, $v \in \R$ it holds
 \begin{equations}\label{dl}
 & \int_\R \left|p(v) -q(v)\right|\, dv =  \int_\R \left|\sqrt{p(v)} -\sqrt{q(v)}\right|\cdot \left|\sqrt{p(v)} +\sqrt{q(v)}\right|\, dv \le \\
 & \left[\int_\R\left(\sqrt{p(v)} -\sqrt{q(v)}\right)^{2}\, dv\right]^{1/2} \cdot \left[\int_\R\left(\sqrt{p(v)} +\sqrt{q(v)}\right)^{2}\, dv\right]^{1/2}  \le \\
 & d_H(p,q)  \left[2\, \int_\R \left({p(v)} +{q(v)}\right)\, dv\right]^{1/2} = \sqrt 2 \,  d_H(p,q) \left( \|p\|_{L_1} +\|q\|_{L_1}\right)^{1/2}.
 \end{equations} 
 Hence, using \fer{bbc} we obtain
  \[
  \int_\R \left|  h(t_n) -  \|h(t_n)\|_{L_1} f_\infty   \right|\, dv \le \frac 2{1-\alpha} d_H( h(t_n), \|h(t_n)\|_{L_1} f_\infty) ,
  \]
 namely the $L_1$-convergence to zero of the sequence $\left\{ h(t_n) -  \|h(t_n)\|_{L_1} f_\infty\right\}_{n \ge 0}$.
 This implies that we can extract from the above sequence of times a subsequence, still denoted by $t_n$, such that on this subsequence 
  \[
   h(v, t_n) -  \|h(t_n)\|_{L_1} f_\infty(v) \to 0 \qquad {\rm a.s.\,\,\,  in}\,\,\, \R_+.
     \]
Since  $f_\infty(v) > 0$, $v \in \R_+$, it holds
 \[
 \frac{f_\infty(v)}{\|h(t_n)\|_{L_1}g(v,t_n)} \to 1\qquad {\rm a.s.\,\,\,  in}\,\,\, \R_+,
 \]
or, what is the same 
 \[
\frac{f_\infty(v)}{\|h(t_n)\|_{L_1}} - g(v,t_n) \to 0 \qquad {\rm a.s.\,\,\,  in}\,\,\, \R_+.
 \]
Integrating on $\R_+$, and recalling that by Theorem \ref{conv-exp}
 \[
 \int_0^\infty g(v,t_n) \, dv \to  \int_0^\infty f_\infty(v) \, dv = 1,
 \]
 shows that 
   \be\label{as}
\lim_{n \to\infty}  \|h(t_n)\|_{L_1} \to 1.
    \ee 
 The validity of \fer{limit6} and \fer{as} then imply 
  \be\label{limit7}
 \lim_{n \to \infty} d_H( h(t_n),  f_\infty) = 0. 
  \ee   
   Indeed
   \begin{equations}
   & d_H( h(t_n),  f_\infty)^2   = d_H( h(t_n), \|h(t_n)\|_{L_1} f_\infty)^2  +\\
   & 1 -  \|h(t_n)\|_{L_1} + 2\, \left( \|h(t_n)\|_{L_1} f_\infty)^{1/2} - 1 \right)\, \int_0^\infty \sqrt{hf_\infty} \, dv,
    \end{equations}
   and
    \[
   \int_0^\infty \sqrt{hf_\infty} \, dv =  \int_0^\infty \sqrt{\frac {f_\infty}{g(t_n)}}f_\infty \, dv  \le \frac 1{\sqrt{1-\alpha}},
    \]
    Last, consider that
    \begin{equations}\label{ddl}
   &d_H(h(t_n),f_\infty)^2 =  \int_0^\infty \left(\frac{f_\infty}{\sqrt{g(t_n)}} - \sqrt{f_\infty}\right)^2\, dv = \\
   & \int_0^\infty \frac{f_\infty}{{g(t_n)}}\left(\sqrt{g(t_n)} - \sqrt{f_\infty}\right)^2\, dv \ge
    \int_{\{ f_\infty \ge g(t_n)\}}\left(\sqrt{g(t_n)} - \sqrt{f_\infty}\right)^2\, dv.
 \end{equations}
    Thanks to \fer{dl}, from \fer{ddl} we obtain
    \be\label{sl}
    d_H(h(t_n),f_\infty) \ge \frac 12 \int_{\{ f_\infty \ge g(t_n)\}}\left(f_\infty -g(t_n)\right)\, dv .
    \ee
    Taking into account that both $f_\infty$ and $g(t_n)$ are probability density functions, it holds
    \be\label{lla}
     \int_{\{ f_\infty \ge g(t_n)\}}\left(f_\infty -g(t_n)\right)\, dv = \frac 12 \| g(t_n)- f_\infty \|_{L_1}.
         \ee
Also, since for $a>b>0$ 
 \[
 (a-b)^2 \le a^2 -b^2,
 \]
one obtains easily the inequality
 \be\label{ldd}
 d_H(g(t_n), f_\infty)^2 \le  \| g(t_n)- f_\infty \|_{L_1} .
 \ee
Grouping all these inequalities we finally get
 \be\label{ fina}
d_H(h(t_n), f_\infty) \ge \frac 14 \| g(t_n)- f_\infty \|_{L_1} \ge \frac 14  d_H(g(t_n), f_\infty)^2.
  \ee   
 It follows that, along the subsequence $\{t_n\}_{n \ge 0} $
  \be\label{gg}
\lim_{n \to \infty} d_H( g(t_n),  f_\infty) = \lim_{n \to \infty}  d_{H,\alpha}(f(t_n),f_\infty) = 0
  \ee   
 However, in view of Theorem \ref{decay-H}, the sequence  $d_{H,\alpha}(f(t),f_\infty)$, $t \ge 0$ is monotonically nonincreasing. This implies that the whole sequence converges to zero as time goes to infinity.

 \medskip

\begin{thm}\label{conv-hel}
Let $f_0(v)$ be a smooth probability density in $\R$ satisfying \fer{n-c},  and such that its moments up to $3+\delta$ are finite for some $\delta >0$. Then,  for  $0<\alpha<1$, the solution to the Fokker--Planck equation \fer{FP2c} converges towards the equilibrium density $f_\infty$  in $\alpha$-Hellinger distance.
\end{thm}

\medskip
 
Theorem \ref{conv-hel} has important consequences.  First, in view of the inequality
 \[
 \| f-g\|_{L_1} \le 2 d_H(f,g),
 \] 
that holds for any pair of probability densities $f,g$, we get, for $0< \alpha <1$
 \[
 (1-\alpha) \int_\R|f(v) -g(v)| \, dv = \int_\R | f - (\alpha f(v)+(1-\alpha)g(v))|\, dv \le   d_{H,\alpha}(f,g).
 \] 
Hence, under the same conditions of Theorem \ref{conv-hel} the convergence to zero in $\alpha$-Hellinger distance implies the $L_1$-convergence of the solution to the Fokker--Planck equation \fer{FP2c} towards its equilibrium density. 

Moreover, as proven in \cite{TT1}, Lemma $\rm 3.3$, the condition of smoothness of the initial value can be dropped as soon as convergence is restricted to $L^1$. 

\medskip
\begin{cor}\label{conv-l1}
Let $f_0(v)$ be a  probability density in $\R$ satisfying \fer{n-c},  and such that its moments up to $3+\delta$ are finite for some $\delta >0$. Then,  the solution to the Fokker--Planck equation \fer{FP2c} converges towards the equilibrium density $f_\infty$  in $L^1$.
\end{cor} 
   
\section{Conclusions}
The Fokker--Planck equation \fer{FP2c} studied in this paper appears as a useful and consistent model to study the evolution in time of the distribution of wealth in a population, even in the realistic case in which part of the agents can have debts. If the total mean wealth of the population is positive, it is shown that the unique equilibrium density, supported in half-line of positive wealths, is still attracting any  density, with part of the mass located  on the negative half-line. At difference with the case studied in \cite{TT1}, where convergence to the equilibrium density has been shown in $L^1$-norm, here convergence with rate has been proven only in terms of a Fourier-based metric, equivalent to the weak$\null^*$-convergence of measures. A rigorous study of the time evolution of relative entropy functionals,  Jensen--Shannon entropy \cite{lin} and $\alpha$-Hellinger distance, shows that  these functionals are monotonically nonincreasing in time, and can be coupled to furnish convergence without rate in $\alpha$-Hellinger distance and consequently in $L_1$. A challenging problem which remains open is to be able to quantify the rate of decay of the solution with respect to the $L_1$-norm.
\vskip 1cm

\section*{Acknowledgement}  This work has been written within the
activities of the National Group of Mathematical Physics (GNFM) of INdAM (National Institute of
High Mathematics), and partially supported by the   MIUR-PRIN Grant 2015PA5MP7 ``Calculus of Variations''.


\end{document}